\documentclass[sigconf]{acmart}

\AtBeginDocument{\providecommand\BibTeX{{\normalfont B\kern-0.5em{\scshape i\kern-0.25em b}\kern-0.8em\TeX}}}

\usepackage[ruled,vlined,linesnumbered]{algorithm2e}

\DeclareMathOperator*{\argmin}{arg\,min}
\usepackage{dsfont,bbm}
\usepackage[mathscr]{eucal}

\usepackage{float}

\usepackage{amsmath}

\usepackage{graphicx}
\usepackage{subcaption}

\def\equationautorefname~#1\null{Eq.~#1\null}
\def\algorithmautorefname~#1\null{Alg.~#1\null}
\usepackage{hyperref}
\usepackage[nameinlink]{cleveref}
\usepackage[english]{babel}
\crefname{subsection}{subsection}{subsections}
\crefname{subsubsection}{subsubsection}{subsubsections}
\crefname{algorithm}{alg.}{algs.}
\crefformat{equation}{#2eq.~#1#3}
\Crefformat{equation}{#2Eq.~#1#3}

\usepackage{letltxmacro}
\LetLtxMacro{\originaleqref}{\eqref}
\renewcommand{\eqref}{Eq.~\originaleqref}

\newtheoremstyle{99problems}{2pt}{2pt}{\itshape}{10pt}{\scshape}{:}{ }{}

\theoremstyle{99problems} \newtheorem{problem}{Problem}
\theoremstyle{99problems} \newtheorem*{problIn}{Input}
\theoremstyle{99problems} \newtheorem*{problOut}{Output}

\newcommand{\pluseq}{\mathrel{+}=}
\newcommand{\lesseq}{\mathrel{-}=}

\copyrightyear{2021}
\acmYear{2021}
\setcopyright{iw3c2w3}
\acmConference[WWW `21]{Proceedings of the Web Conference 2021}{April 19--23, 2021}{Ljubljana, Slovenia}
\acmBooktitle{Proceedings of the Web Conference 2021 (WWW `21), April 19--23, 2021, Ljubljana, Slovenia}
\acmPrice{}
\acmDOI{10.1145/3442381.3450102}
\acmISBN{978-1-4503-8312-7/21/04}

\begin{document}



\author{Giselle Zeno}
\affiliation{\institution{Purdue University}
    \city{West Lafayette}
    \state{IN}
    \country{USA}
}
\email{gzenotor@purdue.edu}

\author{Timothy La Fond}
\affiliation{\institution{Lawrence Livermore National Laboratory}
    \city{Livermore}
    \state{CA}
    \country{USA}
}
\email{lafond1@llnl.gov}

\author{Jennifer Neville}
\affiliation{\institution{Purdue University}
    \city{West Lafayette}
    \state{IN}
    \country{United States}
}
\email{neville@purdue.edu}

\newcommand{\modelname}{DYnamic MOtif-NoDes}
\newcommand{\modelacronym}{DYMOND}
\title{\modelacronym{}: \modelname{} Network Generative Model}

\begin{abstract}
Motifs, which have been established as building blocks for network structure, move beyond pair-wise connections to capture longer-range correlations in connections and activity. In spite of this, there are few generative graph models that consider higher-order network structures and even fewer that focus on using motifs in models of dynamic graphs. Most existing generative models for temporal graphs strictly grow the networks via edge addition, and the models are evaluated using static graph structure metrics---which do not adequately capture the temporal behavior of the network. To address these issues, in this work we propose \modelname{} (\modelacronym{})---a generative model that considers (i) the dynamic changes in overall graph structure using temporal motif activity and (ii) the roles nodes play in motifs (e.g., one node plays the hub role in a wedge, while the remaining two act as spokes). We compare \modelacronym{} to three dynamic graph generative model baselines on real-world networks and show that \modelacronym{} performs better at generating graph structure and node behavior similar to the observed network. We also propose a new methodology to adapt graph structure metrics to better evaluate the temporal aspect of the network. These metrics take into account the changes in overall graph structure and the individual nodes' behavior over time. \end{abstract}

\maketitle

\section{Introduction}\label{Introduction}

Network models provide a way to study complex systems from a wide range of domains, such as social, biological, computing and communication networks. The ability to generate synthetic networks is useful for evaluating systems on a wide(r) range of structure and sharing without divulging private data, for example.

Many generative models for static graphs have aimed to generate synthetic graphs that can simulate real-world networks \cite{chakrabartiGraphMiningLaws2006}. Random graph models \cite{Erdos1959} were among the first proposed. These were adapted to produce degree distributions similar to those of social networks \cite{Wasserman1996, Chung2002}, but still failed to generate the clustering of real-world networks. Block models \cite{Nowicki2001, Leskovec2010, Moreno2009} were later proposed for creating communities observed in social networks. However, these methods focus on capturing either global or local graph properties.

Historically, complex networks with temporal attributes have been studied as static graphs by modeling them as growing networks or aggregating temporal data into one graph. In reality, most of these networks are dynamic in nature and evolve over time, with nodes and edges constantly being added and removed. Initial models for temporal or dynamic networks (where links appear and disappear, such as social-network communication  \cite{Karrer2009, Holme2013, Rocha2013}) focused on modeling the edges over time, ignoring higher-order structures.

Although traditionally most graph models have been edge-based, motifs have been established as building blocks for the structure of networks \cite{Milo2002}. Modeling motifs can help to generate the graph structure seen on real-world networks and capture correlations in node connections and activity. Following work that studied the evolution of graphs using higher-order structures  \cite{Zhao2010, zhang2014dynamic, hulovatyy2015exploring, paranjape2017motifs, Benson2018}, recently \cite{Purohit2018} proposed a generative model using temporal motifs to produce networks where links are aggregated over time. However, this approach assumes that edges will not be removed once they are added (i.e., placed).

In this paper, we propose a dynamic network model, using temporal motifs as building blocks, that generates dynamic graphs with links that appear and disappear over time. Specifically, we propose \modelname{} (\modelacronym), a generative model that first assigns a motif configuration (or motif type) and then samples inter-arrival times for the motifs. One challenge that comes with this is sampling the motif placement. To this end, we define motif node roles and use them to calculate the probability of each motif type. The motifs and node roles can capture correlations in node connections and activity.

Another key challenge is \textit{how to evaluate dynamic graph models?} Previous work has focused on evaluating models using the structure metrics defined for {\em static} graphs without incorporating measures that reflect the temporal behavior of the network. To address this, we adapt previous metrics to consider temporal structure and node behavior over time. Using both sets of metrics (i.e., static and dynamic), we evaluate our proposed model on five real-world datasets, comparing against three recent alternatives. The results show that \modelacronym{} generates dynamic networks with the closest graph structure and similar node behavior as the real-world data.

To summarize, we make the following contributions:
\begin{itemize}
    \item We conduct an empirical study of motif behavior in dynamic networks, which shows that motifs do not change/evolve from one timestep to another, rather they keep re-appearing in the same configuration
    \item Motivated by the above observation, we develop of a novel statistical dynamic-graph generative model that samples graphs with realistic structure and temporal node behavior using motifs
    \item We outline a new methodology for comparing dynamic-graph generative models and measuring how well they capture the underlying graph structure distribution and temporal node behavior of a real graph
\end{itemize}

The rest of the paper is organized as follows. First, we go over related work and discuss where our model fits in \Cref{Related_Work}. Then in \Cref{Motif_Evolution}, we present our empirical study of the evolution of motifs in temporal graphs. In \Cref{Model},
we propose our generative model \modelacronym{}.
In \Cref{Methodology}, we present the datasets and baseline models used in our evaluation. We also describe our evaluation metrics and how to adapt them to dynamic networks.  Finally, we discuss the results of our evaluation (\Cref{Results}) and present our conclusions (\Cref{Conclusion}).  \section{Related Work}\label{Related_Work}

Most models for temporal or dynamic networks have focused on modeling the edges over time \cite{Karrer2009, Holme2013, Rocha2013}. A straightforward approach to generating temporal networks is to generate first a static graph from some model, and for each link generate a sequence of contacts \cite{Holme2015}. Holme \cite{Holme2013} uses an approach where they draw degrees from a probability distribution and match the nodes in random pairs for placing links. Then, for every link, they generate an active interval duration from a truncated power-law distribution and uniform random starting time within that time frame. Rocha and Blondel \cite{Rocha2013} use a similar method where the active interval of a node starts when another node's interval ends.
Another approach is to start with an empty graph. Then, every node is made active according to a probability and connected to $m$ random nodes. Perra et al. \cite{Perra2012} use this approach with a truncated power-law distribution for each node's probability of being active. Laurent et al. \cite{Laurent2015} extend this model to include memory driven interactions and cyclic closure. Other extensions include aging effects \cite{Moinet2015} and lifetimes of links \cite{sunny2015dynamics}. Vestergard et al. \cite{Vestergaard2014} model nodes and links as being active or inactive using temporal memory effects. All of these node and edge-based models do not consider higher-order structures and fail to create enough clustering in the networks generated.

Motivated by the work that established motifs as building blocks for the structure of networks \cite{Milo2002}, the definition of motifs was extended to temporal networks by having all the edges in a given motif occur inside a time period \cite{Zhao2010, hulovatyy2015exploring, paranjape2017motifs}.
Zhang et al. \cite{zhang2014dynamic} study the evolution of motifs in temporal networks by looking at changes in bipartite motifs in subsequent timesteps. Benson et al. \cite{Benson2018} study higher-order networks and how 3-node motifs evolve from being empty to becoming a triangle in aggregated temporal graphs. Purohit et al. \cite{Purohit2018} propose a generative model that creates synthetic temporal networks where links are aggregated over time (i.e., no link deletions).
Zhou et al. \cite{zhouDataDrivenGraphGenerative2020} propose a dynamic graph neural network model that takes into account higher-order structure by using node-biased temporal random walks to learn the network topology and temporal dependencies.
The models that use temporal motifs are not designed for dynamic networks. We propose the first motif-based dynamic network generative model. \section{Motif Evolution}\label{Motif_Evolution}

\begin{figure*}[hbt!]
    \centering
    \begin{subfigure}[b]{0.3\textwidth}
        \includegraphics[trim={.15cm .15cm .15cm 1.6cm},clip,width=1.1\linewidth]{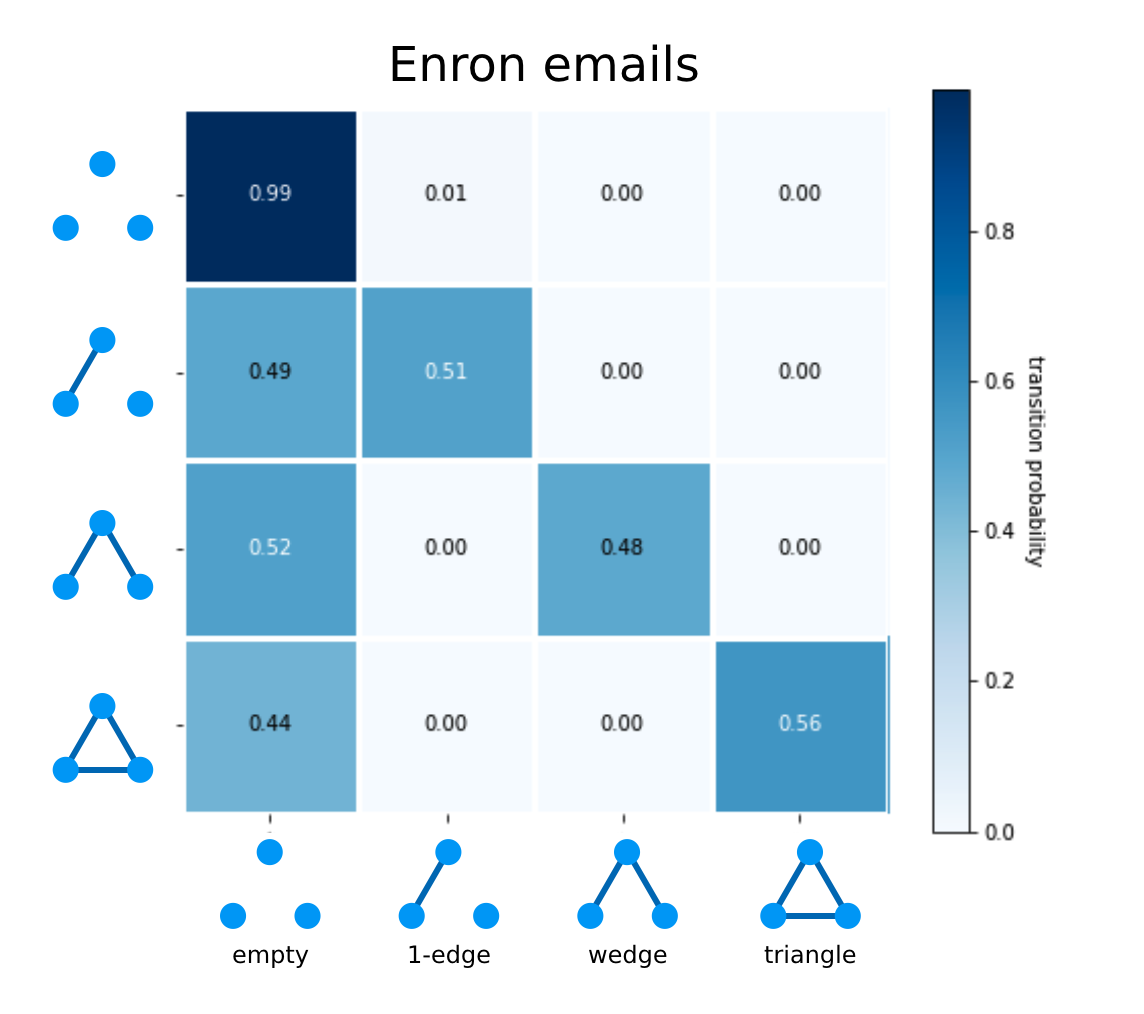}
        \caption{Enron Emails}
        \label{fig:enron_transition_matrix}
        \Description{Transition probability matrix for Enron emails dataset}
    \end{subfigure}
    ~ \begin{subfigure}[b]{0.3\textwidth}
        \includegraphics[trim={.15cm .15cm .15cm 1.6cm},clip,width=1.1\linewidth]{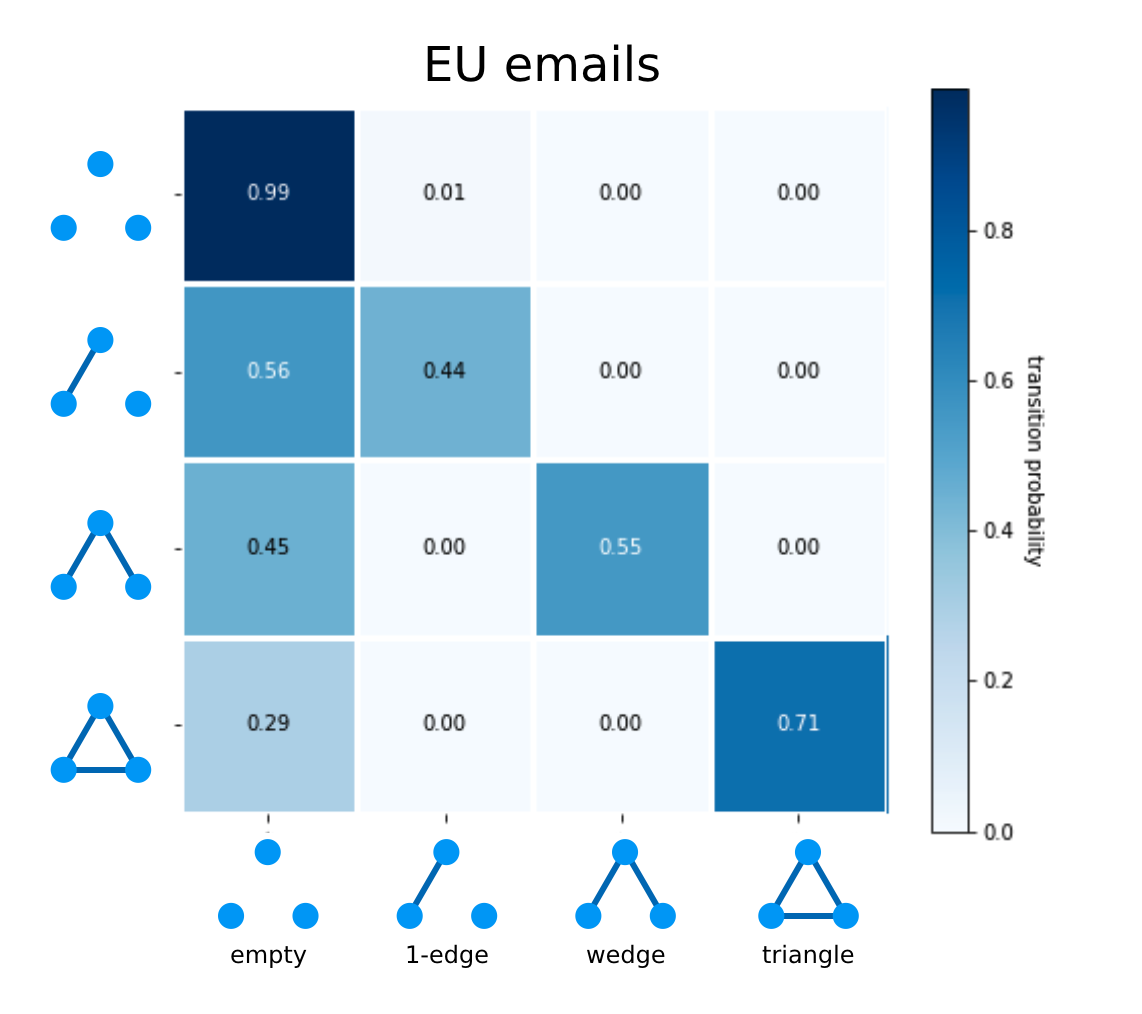}
        \caption{EU Emails}
        \label{fig:eu_transition_matrix}
        \Description{Transition probability matrix for EU emails dataset}
    \end{subfigure}
    ~
    \begin{subfigure}[b]{0.3\textwidth}
        \includegraphics[trim={.15cm .15cm .15cm 1.6cm},clip,width=1.1\linewidth]{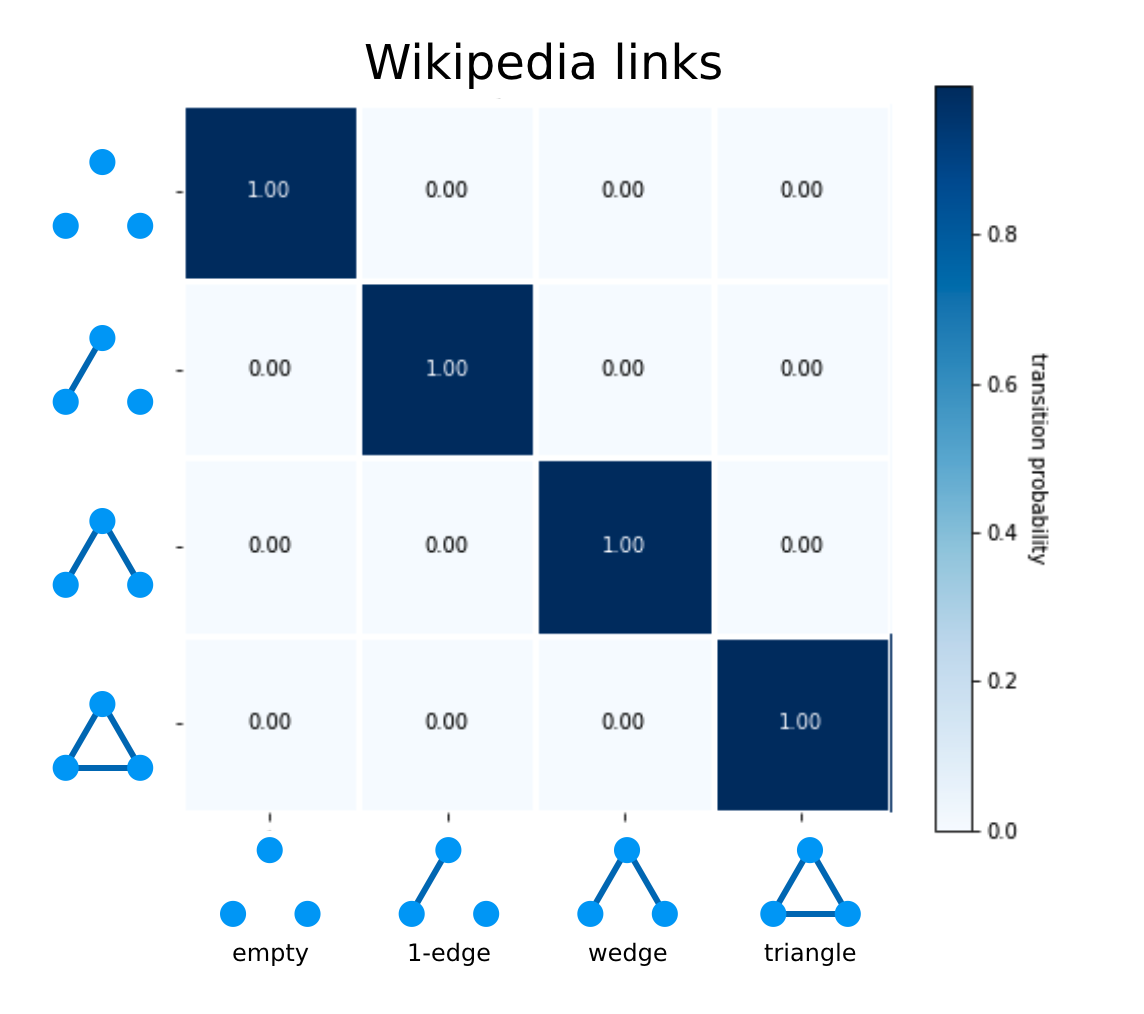}
        \caption{Wikipedia Links}
        \label{fig:wikipedia_transition_matrix}
        \Description{Transition probability matrix for Wikipedia links dataset}
    \end{subfigure}
    \vspace{-3mm}
    \caption{Observed Motif Transition Probabilities}
\end{figure*}

Related work on modeling temporal networks showed the evolution of motifs using a triadic closure mechanism (e.g., wedges becoming triangles) \cite{Benson2018}. However, these works make the assumption that edges will remain in the network once they are added \cite{Benson2018,zhang2014dynamic,Purohit2018}.
This assumption would hold on growing networks, but does not apply to dynamic networks where links can also be removed.

We make the distinction that we are interested in the evolution of motifs in dynamic networks, where edges can appear and disappear. In our initial study below, we investigate if similar motif behavior occurs in dynamic networks across subsequent time windows (e.g., if the motifs appear, merge, split and/or disappear over time). Specifically, we investigate 3-node motifs and look for changes from one motif type to another
(for example, wedges becoming triangles and vice versa).

\subsection{Definitions}\label{Definitions}

Here we introduce our main definitions used in this paper. The rest of notations and symbols are summarized in \autoref{tab:notation}.

\begin{definition}[Graph Snapshot]
A graph snapshot is a time-slice of a network at time $t$, defined as $G_{t} = (V_{t}, E_{t}, S_{t})$, where $V_{t} \subseteq V$ is the set of active nodes, $E_{t} \subseteq E$ is the set of edges at time $t$, and $S_{t} \subseteq S$ are the edge timestamps.
\end{definition}

\begin{definition}[Dynamic Network]
A dynamic network (or graph) $\textbf{G}=\left\{ G_{1}, \ldots ,G_{T}\right\}$ is a sequence of graph time-slices where $T$ is the number of timesteps.
\end{definition}

\begin{definition}[Motif]
We define a motif as a 3-node subgraph $\{u,v,w\}$ and its \textit{motif type} is determined by the number of edges between the nodes (i.e., \textit{empty} has 0, \textit{1-edge} has 1, \textit{wedge} has 2, \textit{triangle} has 3 edges).
\end{definition}

\subsection{Empirical Study}\label{Empirical_Study}

We test the hypothesis that motifs changing configurations is driven by a time-homogeneous Markov process, where the graph structure at the next timestep $t+1$ depends on the current timestep $t$. Each timestep corresponds to a time window of the temporal graph. Then, we consider all $3$-node motifs at each timestep to either transform from one motif type to another or remain the same. Note isomorphisms are combined into the same configuration.

We study the effectiveness of this approach on the Enron Emails and EU Emails datasets, described in
\Cref{dataset_Enron,dataset_EU}
respectively. Additionally, we use a Wikipedia Links dataset, which shows the evolution of hyperlinks between Wikipedia articles \cite{preusse2013structural, Kunegis2013}. The nodes represent articles. Edges include timestamps and indicate that a hyperlink was added or removed depending on the edge weight (-1 for removal and +1 for addition). The transition probability matrices for both email datasets (Enron Emails and EU Emails) show that the motifs with edges (i.e., 1-edge, wedge, and triangle) will either keep their current motif type, or become empty with almost equal probability
(\Cref{fig:enron_transition_matrix,fig:eu_transition_matrix}).
For each motif type with edges, the count of times it stayed is very close to that of becoming empty at the next time period. In contrast, the Wikipedia Links dataset is a growing network, with more links between articles being added and very few removed. This makes it unlikely to see any motif with edges becoming empty (\Cref{fig:wikipedia_transition_matrix}).

In the dynamic network datasets we investigated:
(1) we do not observe motifs with edges changing from one motif type to another (e.g., wedges becoming triangles and vice versa), even when selecting different time windows to create the timesteps, and
(2) motifs stay as the same type or disappear in the next time window. This motivates our use of motifs and inter-arrival rates in our proposed generative model for dynamic networks, which we outline next.

\section{\modelname{} Model}\label{Model}

We formally define the problem of dynamic network generation as follows:
\begin{problem}\textbf{Dynamic Network Generation}
    \begin{problIn}
        A dynamic network $\textbf{G}=\left\{G_{1}, \ldots, G_{T}\right\}$
    \end{problIn}
    \begin{problOut}
        A dynamic network $\textbf{G}'=\left\{G'_{1}, \ldots, G'_{T'}\right\}$,
        where the distribution of graph structure for $\textbf{G}'$ matches $\textbf{G}$
        and node behavior is aligned across $\textbf{G}'$ and  $\textbf{G}$ (i.e., the node behavior of a specific node $v_{i'}$ in $\textbf{G}'$ should be similar to a specific node $v_{i}$ in $\textbf{G}$).
    \end{problOut}
\end{problem}

Specifically, consider an arbitrary graph statistic $s(G)$ (e.g., average path length). Then the distribution of statistic values observed in the input dynamic network $\textbf{s}_{in}=\left\{s(G_{1}), \ldots, s(G_{T})\right\}$ should match the distribution of statistic values observed in the output dynamic network $\textbf{s}_{out}=\left\{s(G'_{1}), \ldots, s(G'_{T'})\right\}$. Similarly, take any node statistic $\textbf{s}(v_i | \textbf{G})$ (e.g., node degree). Then, using
the temporal distribution of values for a node $\textbf{s}(v_i | \textbf{G}) = \{s(v_i | G_1), \ldots, s(v_i | G_T) \}$, the distribution of values for all nodes in the input dynamic network $\{ \textbf{s}(v_j | \textbf{G}) \}_{j \in \textbf{G}}$ should match the distribution of values for all nodes in the output dynamic network $\{ \textbf{s}(v_{j'} | \textbf{G}') \}_{j' \in \textbf{G}'}$.

To generate dynamic networks as specified above, we propose the \modelname{} (\modelacronym{}) model\footnote{Code is available at \url{https://github.com/zeno129/DYMOND}}. Our model
makes the following assumptions about the graph generative process:
\begin{enumerate}
    \item nodes in the graph become active and remain that way,
    \item nodes have a probability distribution over role types that they play in different motifs,
    \item node triplets have a single motif type over time,
    \item there is a distribution of motif types over the set of graphs,
    \item motif occurrences over time are distributed exponentially with varying rate.
\end{enumerate}

First we describe \modelacronym{}'s generative process below. Then we describe our approach to estimate model parameters from an observed dynamic network. We model the time until nodes become active as Exponential random variables with the same rate $\lambda_V$. Since all possible $3$-node motifs are considered, there will be edges shared among them. Therefore, to estimate the inter-arrival rate for each motif, we weigh the count of times a motif appeared by the number of edges shared with other motifs in a timestep. For each motif type with edges (i.e., triangle, wedge, and 1-edge), the model fits an Exponential distribution with the motif inter-arrival rates of that type. Motivated by our findings in \Cref{Motif_Evolution}, when a motif is first sampled it will keep the same configuration in the future.

\begin{figure}[hbt!]
    \centering
    \includegraphics[width=0.8\linewidth]{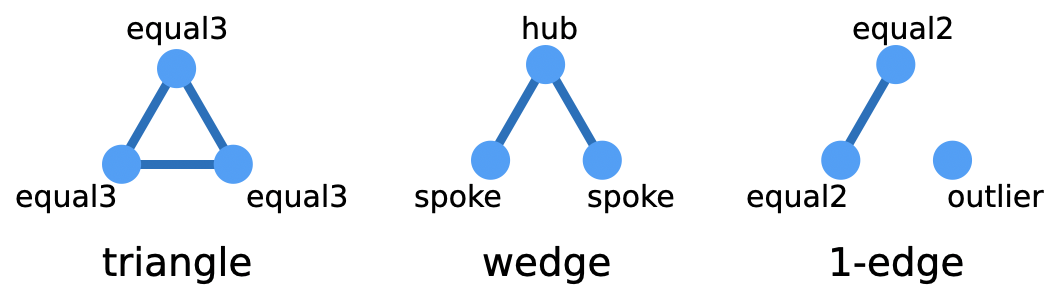}
    \vspace{-3mm}
    \caption{Motif Types and Node Roles}
    \label{fig:motif_node_roles}
    \Description{Motif types and node roles}
\end{figure}

In the generation process, the motifs are sampled from a probability distribution based on the probability of the nodes in a triplet participating in a particular motif type, while also ensuring the motif type proportions in the graph are maintained. The motif type probability for a triplet considers the roles each node would play in a motif. For example, in a wedge one node would be a hub and the other two would be spokes (\Cref{fig:motif_node_roles}). The node role probabilities are learned from the input graph's structure and the motifs that the node participates in.

The motivation for this modeling approach is based on the following conjectures: (1) by modeling higher-order structures (i.e., motifs), the model will capture the underlying distribution of graph structure, and (2) by using the motif roles that nodes take in the dynamic network, the model will also capture correlations in node connections and activity. \subsection{Generative Process}\label{Sampling}

The overall generative process is described in \Cref{GenerateDynamicGraph}: In \cref{alg:SampleDynamicGraph_GetActiveNodes}, we first get the nodes that are active at each timestep using the node arrival rate $\lambda_V$ (see \Cref{GetActiveNodes}, \Cref{Appendix}). Whenever new nodes become active, we calculate the new triplets of active nodes that are now eligible to be sampled as a motif in \cref{alg:SampleDynamicGraph_active_triplets}. In \cref{alg:SampleMotifsWithTimesteps}, we proceed to sample the motifs, based on the node role probabilities $p_R$ for each motif type, and the timesteps the motifs will appear using the motif inter-arrival rates $\lambda_M$ (see \Cref{SampleMotifs}). In \cref{alg:SampleDynamicGraph_SampleMotifEdges}, we place the motifs' edges (\Cref{PlaceMotifEdges}, \Cref{Appendix}) and in \cref{alg:ConstructGraph} we construct the graph (\Cref{ConstructGraph}, \Cref{Appendix}).

\begin{small}
\begin{algorithm}[h]
    \SetKwInput{Input}{input}
    \SetKwInOut{Output}{output}
    \DontPrintSemicolon
    \SetKwFunction{GetActiveNodes}{GetActiveNodes}
    \SetKwFunction{PlaceMotifEdges}{PlaceMotifEdges}
    \SetKwFunction{SampleMotifs}{SampleMotifs}
    \SetKwFunction{ConstructGraph}{ConstructGraph}
    \Input{\hspace{2pt}
        $T$, $N$, $\lambda_V$, $p_M$, $\lambda_M$, $p_R$, $c_R$
    }
    \Output{$\textbf{G}' = \{ G'_{1}, \ldots, G'_{T} \}$}
    \BlankLine
    \Begin{
$V \leftarrow$ \GetActiveNodes{$T,\ N,\ \lambda_V$} \;\label{alg:SampleDynamicGraph_GetActiveNodes}
        $M \leftarrow \emptyset, M^S \leftarrow \emptyset, M^E \leftarrow \emptyset$ \;
        \For{$t \in [1, \ldots, T]$}{
            \tcp{New active triplets at timestep $t$}
            $U_{t} \leftarrow \big\{ m = \{u,v,w\} \subseteq V_{t}\ |\ u<v< w,\ m \notin U_{t-1} \big\}$ \;\label{alg:SampleDynamicGraph_active_triplets}
            $M_{t}, M^T_{t}, M^S_{t}, M^R_{t}, p_R, c_R \leftarrow$ \SampleMotifs{$U_{t}, p_M, \lambda_M$,$p_R, c_R$} \;\label{alg:SampleMotifsWithTimesteps}
            $M \leftarrow M \cup M_{t}$ \tcp*[h]{save new motifs} \;\label{alg:SampleDynamicGraph_set_of_motifs}
            $M^E_{t} \leftarrow$ \PlaceMotifEdges{$M_{t}, M^T, M^R_{t}$} \;\label{alg:SampleDynamicGraph_SampleMotifEdges}
            $M^T \leftarrow M^T \cup M^T_{t}$ \tcp*[h]{store their types} \;
            $M^E \leftarrow M^E \cup M^E_{t}$ \tcp*[h]{store their edges} \;$M^S \leftarrow M^S \cup M^S_{t}$ \tcp*[h]{store their timestamps} \;
        }
        $\textbf{G}' \leftarrow$ \ConstructGraph{$M, M^E, M^S$} \;\label{alg:ConstructGraph}
    }
    \caption{\texttt{GenerateDynamicGraph}}\label{GenerateDynamicGraph}
\end{algorithm}
\end{small}

\begin{small}
\begin{table}[h!]
    \caption{Notations and Symbols}
    \label{tab:notation}
    \vspace{-2mm}
    \begin{tabular}{ll}
        \toprule
        Symbol & Description \\
        \midrule
$\textbf{G}=\left\{G_{1},\ldots, G_{T}\right\}$ & dynamic temporal network \\
        $G_{t} = (V_{t}, E_{t}, S_{t})$ & graph snapshot at time $t$ \\
        $t \in [1,\ldots, T]$ & timestep (time-window) \\
        $T$ & number of timesteps \\
        $N = |V|$ & number of nodes\\
        $V_t \subseteq V$ & set of active nodes at timestep $t$ \\
        $E_{t} \subseteq E$ & set of edges at timestep $t$ \\
        $S_{t}\big((u',v')\big)$ & list of timesteps for edge $(u',v')$ \\
$\lambda_V$ & node arrival rate (see \Cref{eq:node_arrival_rates})\\
        $p_M = (p_M^{(1)}, p_M^{(2)}, p_M^{(3)})$ & motif type proportions (see \Cref{eq:motif_proportions})\\
        $\lambda_M\big(\{u,v,w\}\big)$ & motif inter-arrival rate (see \Cref{eq:motif_interarrival_rate})\\
        $\lambda_M = (\lambda_M^{(1)}, \lambda_M^{(2)}, \lambda_M^{(3)})$ & motif type rates distr. (see \Cref{eq:motif_type_interarrival_rates})\\
        $C^M$ & count times motifs appear (see \Cref{eq:motif_weighted_counts})\\
        $c_t(u',v')$ & number of times $(u',v') \in E_t$ \\
        $r^{(i)}_{t}(u',v')$ & remaining count $(u',v') \in E_t$ \\
        $|N^{(i)}(u',v')|$ & number of motifs type $i$ \\
        $p_T^{(i)}\left( \{u,v,w\} \right)$ & probability of motif $i$ (see \Cref{eq:motif_probabilities})\\
        $p_R$ & node role probabilities (see \Cref{eq:node_role_probabilities})\\
        $c_R$ & node role counts (see output \Cref{GetNodeRoleCounts})\\
$M$ & set of motifs \\
        $M^{(i)}$ & motifs of type $i$ \\
$M^T$ & motif types \\
        $M^E$ & motif edges \\
        $M^S$ & motif timesteps \\
        $M^R$ & motif node roles \\
\bottomrule
    \end{tabular}
\end{table}
\end{small}
In \Cref{SampleMotifs} \cref{alg:sample_expected_count}, the model calculates the expected count of motifs $n^{(i)}$ to be sampled for each motif type $i$ using the motif type proportions $p_M$. Then in \cref{alg:sample_motif_types_using_node_roles}, the expected number of motifs for each type is sampled, given the probability $p_T$ that the nodes in the triplet take on the roles needed (\Cref{eq:motif_probabilities}). Each node will have an expected count $c_R$ of times they will appear having each role, for the total number of timesteps $T$ to be generated. For this reason, in \cref{alg:sample_motif_timesteps} we sample the timesteps each motif will appear in (\Cref{SampleMotifTimesteps}, \Cref{Appendix}), and in \cref{alg:sample_node_roles} we use the timestep counts to sample the node roles (\Cref{SampleNodeRoles}).

\begin{small}
\begin{algorithm}[h]
    \SetKwInput{Input}{input}
    \SetKwInOut{Output}{output}
    \DontPrintSemicolon
    \Input{\hspace{2pt}
        $U_{t}$, $p_M$, $\lambda_M$, $p_R$, $c_R$
    }
    \Output{$M_t\ $ \texttt{// sampled motifs}\\
        $M^T_{t}$ \texttt{// motif types}\\
        $M^S_{t}$ \texttt{// motif timestamps}\\
        $M^R_{t}$ \texttt{// motif node roles}\\
        $p_R\ $  \texttt{// node role probabilities}\\
        $c_R\ $  \texttt{// node role counts}
    }
    \SetKwFunction{SampleMotifTimesteps}{SampleMotifTimesteps}
    \SetKwFunction{SampleNodeRoles}{SampleNodeRoles}
    \SetKwFunction{UpdateRoleProbs}{UpdateRoleProbs}
    \Begin{
        $L = U_{t}$ \tcp*[h]{triplets to sample from} \;
        \For{$i \in [3,2,1]$}{
            \tcp{sample motifs}
            $n^{(i)} = |U_{t}| \cdot p_M^{(i)}$ \tcp*[h]{num. motifs to sample} \;\label{alg:sample_expected_count}
$M^{(i)} \sim Mult\left(\left[ \frac{p_T^{(i)}\left( \{u,v,w\}\right)}{\sum_{\{u', v', w'\} \in L}{p_T^{(i)}\left( \{u',v',w'\} \right)}}\ \Big|\ \{u,v,w\} \in L \right],\ n^{(i)}\right)$ \;\label{alg:sample_motif_types_using_node_roles}
            $M^T_t = i,\ \forall \{u,v,w\} \in M^{(i)}$ \tcp*[h]{store motif types} \;
            $L = L - M^{(i)}$ \tcp*[h]{triplets left to sample from} \;
            \BlankLine
            $S^{(i)} \leftarrow $ \SampleMotifTimesteps{$t, M^{(i)}, i, \lambda_M$} \;\label{alg:sample_motif_timesteps}
            $M^{(i)'} = \left\{ m\ \big|\ m \in M^{(i)},\ |S^{(i)}(m)| > 0 \right\}$ \;
            $R^{(i)}, p_R, c_R \leftarrow $ \SampleNodeRoles{$M^{(i)'}, i, p_R, c_R, S^{(i)}$} \;\label{alg:sample_node_roles}
            $M^S_{t} \leftarrow M^S_{t} \cup S^{(i)}$ \tcp*[h]{store timesteps} \;
            $M^R_{t} \leftarrow M^R_{t} \cup R^{(i)}$ \tcp*[h]{store roles} \;
        }

    }
    \caption{\texttt{SampleMotifs}}\label{SampleMotifs}
\end{algorithm}
\end{small}
\begin{small}
\begin{algorithm}[hbt!]
    \SetKwInput{Input}{input}
    \SetKwInOut{Output}{output}
    \DontPrintSemicolon
    \Input{\hspace{2pt}
        $M^{(i)'}$, $i$, $p_R$, $S^{(i)}$
    }
    \Output{$R^{(i)}$ \texttt{// node roles for motifs in} $M^{(i)}$\\
        $p_R\ \ $ \texttt{// updated role probabilities}\\
        $c_R\ \ $ \texttt{// updated role counts}
    }
    \Begin{
        \For{$m \in M^{(i)'}$}{
            \If(\tcp*[h]{triangle}){$i = 3$}{
                $R^{(i)}(m, \texttt{equal3}) \leftarrow m $  \;
                $c_R(v,\texttt{equal3}) \lesseq |S^{(i)}(m)|,\ \forall v \in m$ \;
            }
            \ElseIf(\tcp*[h]{wedge}){$i = 2$}{
                \tcp{hub node}
                $p_h = \left[ \frac{ p_R(v,\ \texttt{hub}) }{ \sum_{v' \in m}{p_R(v',\ \texttt{hub})} }\ \Big|\ v \in m \right]$  \;
                $v_h \sim Bin(m,\ p_h)$ \;
                $R^{(i)}(m, \texttt{hub}) \leftarrow v_h$ \;
                $c_R(v_h,\texttt{hub}) \lesseq |S^{(i)}(m)|$ \;
                \tcp{spoke nodes}
                $R^{(i)}(m, \texttt{spoke}) \leftarrow m \setminus \{v_h\}$ \;
                $c_R(v,\texttt{spoke}) \lesseq |S^{(i)}(m)|,\ \forall \{v \in m: v \neq v_h\}$ \;
            }
            \ElseIf(\tcp*[h]{1-edge}){$i = 1$}{
                \tcp{outlier node}
                $p_o = \left[ \frac{ p_R(v,\ \texttt{outlier}) }{ \sum_{v' \in m}{p_R(v',\ \texttt{outlier})} }\ \Big|\ v \in m \right]$  \;
                $v_o \sim Bin(m,\ p_o)$ \;
                $R^{(i)}(m, \texttt{outlier}) \leftarrow v_o$ \;
                $c_R(v_o,\texttt{outlier}) \lesseq |S^{(i)}(m)|$ \;
                \tcp{equal2 nodes}
                $R^{(i)}(m, \texttt{equal2}) \leftarrow m \setminus \{v_o\}$ \;
                $c_R(v,\texttt{equal2}) \lesseq |S^{(i)}(m)|,\ \forall \{v \in m: v \neq v_o\}$ \;
            }
            \For(\tcp*[h]{update role distr (\Cref{eq:node_role_probabilities})}){$v \in m$}{
                $p_R(v,\texttt{role}) = P[v\text{ is }\texttt{role}],\ \forall r \in R$ \;
            }
        }
    }
    \caption{\texttt{SampleNodeRoles}}\label{SampleNodeRoles}
\end{algorithm}
\end{small} \subsection{Learning}\label{Learning}

Given an observed dynamic graph $\textbf{G}$, we estimate the input parameters for our generative process as outlined in \Cref{LearnParameters}, \Cref{Appendix}.

\subsubsection{Node Arrivals}
We begin by estimating the node arrival rate $\widehat{\lambda}_V$, which will determine when nodes become active in the dynamic network, by using the first timestep in which each node had its first edge.
\begin{equation}\label{eq:node_arrival_rates}
\widehat{\lambda}_{V} = \frac{ \sum_{v \in V}{\big( \argmin_t \mathds{1}(v \in V_t) \big)} }{|V|}
\end{equation}

\subsubsection{Motif Proportions}

In \Cref{GetMotifsGraph} (\Cref{Appendix}), we find the motifs in $G_t$ for each time window $t$. For each $3$-node motif $\{u,v,w\}$, we find its motif type $i$ at timestep $t$ (\cref{alg:find_motif_type}). If we have previously seen $\{u,v,w\}$ and the motif type $i$ is of higher order at the current timestep $t$, then we update the type stored to be $i$ (\cref{alg:prefer_higher_order}). For example, if we observe the triplet $\{u,v,w\}$ is a triangle at timestep $t$ and we previously saw it as a wedge, we update $M^T\big( \{u,v,w\} \big)$ as a triangle.

Then, we calculate the motif proportions $\widehat{p}_M^{(i)}$ of each type in the graph, where $i$ corresponds to the number of edges in the motif (i.e., $i=1$ for a 1-edge, $i=2$ for a wedge, and $i=3$ for a triangle motif).
\begin{align}\label{eq:motif_proportions}
    \widehat{p}_M^{(i)} &= \frac{ \big\{ \{u,v,w\} \in M\ \big|\ M^T\big(\{u,v,w\}\big) = i \big\} }{\binom{N}{3}}, & \text{for }i \in [1, 2, 3] \nonumber\\
        \widehat{p}_M^{(0)} &= 1 - \sum_{i=1}^{n}{\widehat{p}_M^{(i)}}
\end{align}
where $M$ is the set of motifs, and $\{u,v,w\}$ is a motif consisting of the nodes $u$, $v$, $w$.

\subsubsection{Motif Inter-Arrivals}

We estimate the inter-arrival rates of each observed motif $ \{u,v,w\}$ using weighted edge counts (\Cref{eq:motif_interarrival_rate}). Their rates are then used to learn a rate of inter-arrival rates $\widehat{\lambda}_{M}^{(i)}$ from the motifs of each type $i$ (\Cref{eq:motif_type_interarrival_rates}). Note that we do not need to estimate rates for the empty motif type ($i=0$).
\begin{subequations}
    \begin{equation}\label{eq:motif_interarrival_rate}
        \widehat{\lambda}_M\big(\{u,v,w\}\big) = \frac{ \sum_{t=1}^{T}{C_{t}^{M}\big(\{u,v,w\}\big)} }{T}
    \end{equation}
    \begin{equation}\label{eq:motif_type_interarrival_rates}
        \widehat{\lambda}_{M}^{(i)} = \frac{ \sum_{\{u,v,w\} \in M^{(i)}}{\Big( \widehat{\lambda}_M\big(\{u,v,w\}\big) \Big)} }{|M^{(i)}|}
    \end{equation}
\end{subequations}
where $M^{(i)}$ is the set of all motifs of type $i$.

Since edges might be shared by more than one motif, we use edge-weighted Poisson counts $C_{t}^{M}$, per timestep $t$, to estimate the inter-arrival rate for each motif $\{u,v,w\}$ (\Cref{eq:motif_weighted_counts}). The weights $W_{t}^{(i)}$ will depend on the motif type $i$ of $\{u,v,w\}$ and are calculated for each edge of the motif (\Cref{eq:weighted_counts}).
\begin{equation}\label{eq:motif_weighted_counts}
    C_{t}^{M}\big(\{u,v,w\}\big) = \frac{\sum_{(u',v') \in E_{t}\big(\{u,v,w\}\big)} {W_{t}^{(i)}\big((u',v')\big)}}{\big|E_{t}(\{u,v,w\})\big|}
\end{equation}

For a motif $\{u,v,w\}$, we calculate the weight of its edge $(u',v')$ using the count for the edge in the timestep window and considering its motif type $i$ (\Cref{eq:weighted_counts}). We give larger edge-weight to motif types with more edges, since they are more likely to produce the observed edges.
This also ensures that motif types with smaller proportion ${p}_M^{(i)}$ (\Cref{eq:motif_proportions}) have a high enough inter-arrival rate to show up (i.e., triangles).
\allowdisplaybreaks{
\begin{subequations}\label{eq:weighted_counts}
    \begin{align}
    W^{(i)}_{t}(u',v') &= \frac{r^{(i)}_{t}(u',v')}{|N^{(i)}(u',v')|}\label{eq:weights_wedge_1edge}\\
    r^{(i-1)}_{t}(u',v') &= \min \left(r^{(i)}_{t}(u',v'),\ |N^{(i-1)}(u',v')| \right) \label{eq:remaining_count}
    \end{align}
\end{subequations}}
where $|N^{(i)}(u',v')|$ is the number of motifs of type $i$, the number of times $(u',v')$ appears in $E_t$ is $c_t(u',v')$, the remaining edge count is $r^{(i)}_{t}(u',v')$, and for triangles $r^{(i+1)}_{t}(u',v') = c_t(u',v')$.

\subsubsection{Motif Types}

The probability of a node triplet becoming a triangle, wedge, or 1-edge motif is based on the probability that each node takes on the roles needed to form that motif type. The roles for each motif type are shown in \Cref{fig:motif_node_roles}. Specifically, a triangle requires all three nodes to have the \texttt{equal3} role, a wedge requires one node to be a \texttt{hub} and the rest to have the \texttt{spoke} role, a 1-edge requires two nodes to have the \texttt{equal2} role and the remaining one the \texttt{outlier} role (\Cref{eq:motif_probabilities}).
\begin{equation}\label{eq:motif_probabilities}
    p_{T}^{(i)}\big(\{u,v,w\}\big) = P[u\text{ is }r_1 \land v\text{ is }r_2 \land w\text{ is }r_2]
\end{equation}

\begin{equation}\label{eq:node_role_probabilities}
        P[u\text{ is }r] = \frac{count(u, r)}{\sum_{r' \in R}{count(u, r')}}
\end{equation}
where $R = \{\texttt{equal3}, \texttt{hub}, \texttt{spoke}, \texttt{equal2}, \texttt{outlier}\}$ is the set of possible roles,
and $count(u, r)$ is the weighted count of times that node $u$ had role $r$
(see \Cref{GetNodeRoleCounts}, \Cref{Appendix}). The weights are used to avoid over-counting the roles for motifs of the same type with a shared edge.

\section{Methodology}\label{Methodology}

We first describe the baseline models (\Cref{Baselines}) and datasets (\Cref{Datasets}) used in our evaluation. Then, we introduce the  metrics for evaluating graph structure, our novel approach for evaluating node behavior, and implementation details of all models (\Cref{Evaluation}).

\subsection{Baselines}\label{Baselines}

The related work using motif-based models for temporal graphs focuses on the aggregated temporal graph and not its dynamic changes over time \cite{Purohit2018}. With that in mind, we picked baselines that aim to model the changes in dynamic graphs.
We compare our model with three baselines: a temporal edge-based model (SNLD), a model based on node-activity (ADN), and a graph neural network (GNN) model based on temporal random walks (TagGen).

\subsubsection{Static Networks with Link Dynamics Model (SNLD)}\label{baseline_SNLD}
We used an approach based on \cite{Holme2013}, where they begin by generating a static graph and then generate a series of events. Their procedure begins by sampling degrees from a probability distribution. They refer to these degrees as ``stubs'' and they create links by connecting these ``stubs'' randomly. Finally, for each link, they assign a time-series from an inter-event distribution.

In our implementation of the SNLD model, we start by sampling the degrees from a Truncated Power-law distribution. Since our starting point is a static graph, we assume all the nodes to be active already. Then, we sample inter-event times for every edge. We found that we could best model the edge inter-event times in the real data using an Exponential distribution. To learn the Truncated Power-law parameters, we aggregated and simplified the real graph.

\subsubsection{Activity-Driven Network Model (ADN)}\label{baseline_ADN}
We use the approach in \cite{Laurent2015}, which extends the model in \cite{Perra2012} by adding memory effects and triadic closure. The triadic closure takes place when node $i$ connects to node $k$ forming a triangle with its current neighbor $j$. Adding a triadic closure mechanism helps to create clustering (communities) \cite{bianconiTriadicClosureBasic2014}. The memory effect is added by counting the number of times that the nodes have connected up to the current time $t$.
The procedure starts by creating an empty graph $G_t$ at each timestep. Then, for each node $i$: delete it with probability $p_d$ or mark it as active with probability $a_i$. If the node is ``deleted'', then the edges in the current timestep are removed, the counts of connections set to zero, and another degree is sampled to estimate a new $a_i$. If a node $i$ is sampled as active, we connect it to either: (1) a neighbor $j$, (2) a neighbor of $j$, or (3) a random node.

In our implementation of the SNLD model, we base the probability to create new edges $a_i$ on the degree of node $i$, which we sample from a Truncated Power-law distribution. We estimate the parameters using the average degree across timesteps for the nodes in the real graph. There is a fixed probability $p_d$ for any node being ``deleted'' (losing its memory of previous connections and sampling a new $a_i$). We estimate this probability using the average ratio of nodes becoming disconnected in the next timestep. To estimate the probability for triadic closure (forming a triangle), we use the average global clustering coefficient across timesteps.

\subsubsection{TagGen}\label{baseline_TagGen} TagGen is a deep graph generative model for dynamic networks \cite{zhouDataDrivenGraphGenerative2020}. In their learning process they treat the data as a temporal interaction network, where the network is represented as a collection of temporal edges and each node is associated with multiple timestamped edges at different timestamps. It trains a bi-level self-attention mechanism with local operations (addition and deletions of temporal edges), to model and generate synthetic temporal random walks for assembling temporal interaction networks. Lastly, a discriminator selects generated temporal random walks that are plausible in the input data, and feeds them into an assembling module. We used the available implementation of TagGen\footnote{\url{https://github.com/davidchouzdw/TagGen}} to learn the parameters from the input graph and assemble the dynamic network using the generated temporal walks. \begin{figure*}[h]
    \centering
\begin{subfigure}[b]{0.40\textwidth}
        \includegraphics[trim={.15cm .15cm .15cm 1.1cm},clip,width=\linewidth]{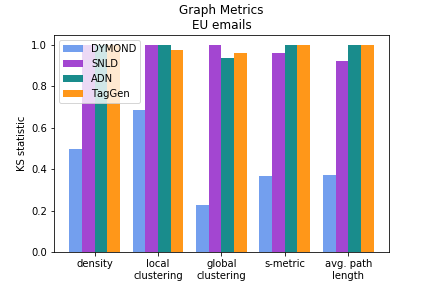}
        \caption{Graph Structure Metrics}
        \label{fig:dataset_graph_metrics-EU}
        \Description{EU Emails - Graph Structure Metrics}
    \end{subfigure}
    ~ \begin{subfigure}[b]{0.40\textwidth}
        \includegraphics[trim={.15cm .15cm .15cm 1.1cm},clip,width=\linewidth]{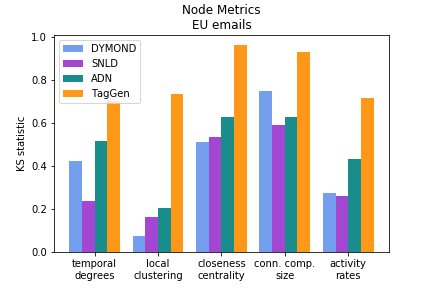}
        \caption{Node Behavior Metrics}
        \label{fig:dataset_node_metrics-EU}
        \Description{EU Emails - Node Behavior Metrics}
    \end{subfigure}
    \vspace{-3mm}
    \caption{EU Emails - KS Statistics}
    \label{fig:dataset_metrics_example}
\end{figure*}
 \subsection{Datasets}\label{Datasets}

We use the datasets described below, with more detailed statistics shown in \Cref{tab:dataset_statistics} and \Cref{fig:dataset_graph_metrics} of \Cref{Appendix}.

\subsubsection{Enron Emails}\label{dataset_Enron}
The Enron dataset is a network of emails sent between employees of Enron Corporation \cite{klimt2004enron, Kunegis2013}. Nodes in the network are individual employees and edges are individual emails. Since it is possible to send an email to oneself, loops were removed.

\subsubsection{EU Emails}\label{dataset_EU}
The EU dataset is an email communication network of a large, undisclosed European institution \cite{leskovec2007graph, Kunegis2013}. Nodes represent individual persons and edges indicate at least one email has been sent from one person to the other. All edges are simple and spam emails have been removed from the dataset.

\subsubsection{DNC Emails}\label{dataset_DNC}
The DNC dataset is the network of emails of the Democratic National Committee that were leaked in 2016 \cite{pickhardtDNC2016, Kunegis2013}. The Democratic National Committee (DNC) is the formal governing body for the United States Democratic Party. Nodes in the network correspond to persons and an edge denotes an email between two people. Since an email can have any number of recipients, a single email is mapped to multiple edges in this dataset.

\subsubsection{Facebook Wall-Posts}\label{dataset_Facebook}

The Facebook dataset is a network of a small subset of posts to other users' walls on Facebook \cite{viswanathEvolutionUserInteraction2009, Kunegis2013}. The nodes of the network are Facebook users, and each directed edge represents one post, linking the users writing a post to the users whose wall the post is written on. Since users may write multiple posts on a wall, the network allows multiple edges connecting a single node pair. Since users may write on their own wall, loops were removed.

\subsubsection{CollegeMsg}\label{dataset_CollegeMsg}

The CollegeMsg dataset is comprised of private messages sent on an online social network at the University of California, Irvine \cite{panzarasaPatternsDynamicsUsers2009, snapnets}. Users could search for other users in the network, based on profile information, and then begin conversation. An edge $(j, k, t)$ means that user $j$ sent a private message to user $k$ at time $t$. \subsection{Evaluation}\label{Evaluation}

We use two sets of metrics in our evaluation for graph structure and node behavior. The majority of graph structure metrics we selected are widely used to characterize graphs. With these first set of metrics we aim to measure if the overall graph structure of the generated graph $\textbf{G}'$ is similar to the dataset graph $\textbf{G}$. For the second set, we propose to use node-aligned metrics to capture node behavior.

\subsubsection{Graph Structure Metrics}\label{Metrics_Graph}
We use the following graph metrics: density, average local clustering coefficient, global clustering coefficient, average path length of largest connected component (LCC), and s-metric. \textit{Density} measures ratio of edges in the graph versus the number of edges if it were a complete graph. The \textit{local clustering coefficient} quantifies the tendency of the nodes of a graph to cluster together, and the \textit{global clustering coefficient} measures the ratio of closed triplets (triangles) to open and closed triplets (wedges and triangles). The \textit{average (shortest) path length}, for all possible pairs of nodes, measures the efficiency of information transport. The \textit{s-metric}, which is less well-known, measures the extent to which a graph has hub-like structure \cite{Li2006}. The s-metric reflects large star structures in a graph. Together with local and global clustering, these metrics provide insight into graph structure like tightly knit-groups and large star structures.

To compare the graph structure generated by the models against that of the datasets, we calculate the graph structure metrics for each time-slice of the generated graph and the input graph. Specifically, for each graph structure metric $s$, we calculate the distribution of values $\textbf{s}_{gen}$ of the generated graph and $\textbf{s}_{in}$ of the input graph (where $\textbf{s}_{in}=\left\{s(G_{1}), \ldots, s(G_{T})\right\}$, $G_{t}$ is a time-slice of $\textbf{G}$, and $t \in [1,\ldots,T]$). Given that we aim to model the distribution of graph structure, and not just generate the same graph sequence, we calculate the Kolmogorov-Smirnov (KS) test statistic on $\textbf{s}_{gen}$ and $\textbf{s}_{in}$ to evaluate $\textbf{G}'$ against $\textbf{G}$.

\subsubsection{Node Behavior Metrics}\label{Metrics_Node}
We propose a new approach to analyze a node's behavior over time.
For this, we use the following {\em node-aligned, temporal} metrics: activity rate, temporal degree distribution, clustering coefficient, closeness centrality, and the size of its connected component. The \textit{activity rate} of a node measures how often it participates in an edge. The \textit{temporal degree distribution} of a node $u$ is the set of degrees of $u$ over all snapshots $G_t$; it shows how many nodes it interacts with. The \textit{local clustering coefficient} of a node measures how close its neighbors are to becoming a clique. The \textit{closeness centrality} of a node $u$ in a (possibly) disconnected graph is the sum of the reciprocal of shortest path distances to $u$ over all other reachable nodes. The node's closeness centrality and size of its connected component indicate the location of the node relative to others.

To compare the temporal node behavior of the generated graphs against the datasets, we calculate the node-aligned temporal metrics for every node in the dataset $\textbf{G}$ and in the generated graphs $\textbf{G}'$. Node-alignment refers to assumption that node ids are aligned over graph snapshots, within a graph sequence. Based on this, we can measure the distribution of values a node has over time $\textbf{s}(v_i | \textbf{G}) = \{s(v_i | G_1), \ldots, s(v_i | G_T) \}$ for each metric $s$. Since the nodes in $\textbf{G}$ do not necessarily correspond to those in $\textbf{G}'$, we consider the inter-quartile range (IQR) of values over time $\{ \textbf{s}(v_j | \textbf{G}) \}_{j \in \textbf{G}}$. We then perform a 2-dimensional KS test using the $Q_1$ and $Q_3$ values of all nodes in $\textbf{G}$ and $\textbf{G}'$. In this way, we capture each node's individual behavior and their joint behavior.

In the past, people have used the mean and median of the values, but these statistics do not capture characteristics of the distribution of values and can be misleading. For example, a synthetic graph $\textbf{G}'$ could have mean and median values of a metric $s$ very close to those of an observed graph $\textbf{G}$, but have a much larger dispersion of $s$ values than observed in $\textbf{G}$.
We use the KS test on the inter-quartile range (i.e., $Q_1$ and $Q_3$) because it does not make assumptions about the distribution of values and can capture variability or dispersion.

\subsubsection{Implementation Details}\label{Implementation}
We estimate initial motif configurations and all parameters of our \modelacronym{} model as described in \Cref{Learning} from dataset graphs. Similarly, we estimate all parameters of the SNLD and ADN baselines as described in \Cref{baseline_SNLD,baseline_ADN}. For the TagGen baseline, we use the available implementation from \cite{zhouDataDrivenGraphGenerative2020}, which is generally described in \Cref{baseline_TagGen}.
  \section{Results}\label{Results}

\subsection{Evaluation of Generated Graphs}
In \Cref{fig:dataset_metrics_example}, we show the KS statistic (lower is better) for the graph structure and node behavior of the EU Emails dataset, as an example. The full set of results of all the other datasets, for the five graph metrics and the five node metrics, are in \Cref{Appendix}. In order to compare the models more easily, we calculated the mean reciprocal rank (MRR) of the KS statistic for the graph structure and the node behavior metrics. To calculate the MRR, we ranked the model results by using the average KS statistic.

In \Cref{tab:map_graph}, we can observe that our \modelacronym{} model outperforms the baselines when considering all the graph structure metrics together using the MRR (higher is better). In \Cref{tab:map_node}, our model performs best on the node behavior for two of the datasets (Enron Emails and Facebook). SNLD performs better on the EU Emails dataset, with our model being a close second, and the CollegeMsg dataset. Finally, ADN performs best on the DNC Emails dataset, but \modelacronym{} significantly outperforms the other two baselines.

\begin{small}
\begin{table}[h!]
    \caption{Graph Structure Mean Reciprocal Rank}
    \label{tab:map_graph}
\begin{tabular}{lrrrrr}
        \toprule
        Model & Enron & EU & DNC & Facebook & CollegeMsg \\
        \midrule
        \modelacronym{} & \textbf{0.57} & \textbf{1.00} & \textbf{0.80} & \textbf{0.77} & \textbf{0.90} \\
        SNLD & 0.52 & 0.61 & 0.29 & 0.47 & 0.45 \\
        ADN & 0.46 & 0.67 & 0.34 & 0.50 & 0.43 \\
        TagGen & 0.53 & 0.58 & 0.64 & 0.30 & 0.30 \\
        \bottomrule
    \end{tabular}
\end{table}

\begin{table}[h!]
    \caption{Node Behavior Metrics Mean Reciprocal Rank}
    \label{tab:map_node}
\begin{tabular}{lrrrrr}
        \toprule
        Model & Enron & EU & DNC & Facebook & CollegeMsg \\
        \midrule
        \modelacronym{} & \textbf{0.93} & 0.70 & 0.50 & \textbf{0.85} & 0.48 \\
        SNLD & 0.40 & \textbf{0.73} & 0.35 & 0.65 & \textbf{0.95} \\
        ADN & 0.47 & 0.37 & \textbf{0.93} & 0.42 & 0.48 \\
        TagGen & 0.25 & 0.25 & 0.27 & 0.25 & 0.25 \\
        \bottomrule
    \end{tabular}
\end{table}
\end{small}

\subsection{Discussion}

SNLD creates a static graph with a degree distribution learned from the input graph and models the edge inter-event times independently. This fails to create graph structure similar to the datasets due to little clustering. The CollegeMsg dataset has low clustering (local and global) but a high s-metric, which indicates large star structure in the graph (i.e., high degree nodes), as seen in \Cref{fig:dataset_graph_metrics}. In this case, SNLD is able to better match the clustering than the other datasets (\Cref{fig:dataset_node_metrics-CollegeMsg}).

ADN models node activation rates using sampled node degrees from a Power-law distribution. However, during sampling a node might be ``deleted'' and have its rate changed. When evaluating node-aligned metrics over time, these rate modifications will change the behavior of a node. This explains the poor performance of ADN on the node activity rates metric of all the datasets  (\Cref{fig:dataset_node_metrics-Enron,fig:dataset_node_metrics-DNC,fig:dataset_node_metrics-Facebook,fig:dataset_node_metrics-CollegeMsg}) except the EU Emails dataset (\Cref{fig:dataset_node_metrics-EU}). Lastly, this model incorporates a triadic closure mechanism to create clustering in the graph structure, which actually helps it perform better on datasets with high clustering (\Cref{fig:dataset_node_metrics-DNC}).

Though TagGen doesn't perform well on most of the graph structure metrics, it manages to create graph clustering comparable to our model in three of the datasets: Enron Emails, DNC Emails, and Facebook (\Cref{fig:dataset_graph_metrics-Enron,fig:dataset_graph_metrics-DNC,fig:dataset_graph_metrics-Facebook}). These three datasets have various star structures (very high degree nodes) across timesteps and shorter diameter than the other datasets. TagGen performs biased temporal random walks and, according to the authors, high degree nodes tend to be highly active resulting in a weak dependence between the topology and temporal neighborhoods. This would explain why it performs better in these three datasets than the others.

Using motifs helps our \modelacronym{} model create better clustering in the graph than the other baselines (\Cref{fig:dataset_graph_metrics-Enron,fig:dataset_graph_metrics-Facebook,fig:dataset_graph_metrics-CollegeMsg}), which also impacts the graph density and s-metric that is generated. The motif inter-arrival times are able to capture the node-aligned behavior in the graph  (\Cref{fig:dataset_node_metrics-Enron,fig:dataset_node_metrics-DNC,fig:dataset_node_metrics-Facebook}).
The addition of motif node roles determines the placement of the motifs in the graph, which in turn impacts the node closeness and shortest path lengths produced. These roles also help capture the node-aligned temporal degree distribution even though we do not directly optimize it.

\section{Conclusion}\label{Conclusion}

Our proposed dynamic network generative model, \modelname{} (\modelacronym{}), is the first motif-based dynamic network generative model. \modelacronym{} not only considers the dynamic changes in overall graph structure using temporal motif activity, but also considers the roles the nodes play in motifs (e.g., one node plays the hub role in a wedge, while the remaining two act as spokes). We note that using motifs helps our \modelacronym{} model create better graph structure overall, while the motif node roles can better represent the temporal node behavior.

In our empirical study of dynamic networks, we demonstrated that motifs with edges: (1) generally do not change configurations (e.g., wedges becoming triangles and vice versa); (2) once they appear, they will continue during the next time window or disappear. Though we do not explicitly address higher-order motifs, using the node roles distribution with graphlets of size three captures some of the dependencies beyond pairwise links. We highlight that we consider all possible motifs from induced 3-node graphlets (i.e., there are overlaps). It is not clear that using higher-order than size three is necessary, but it should be a relatively simple extension to our model.

We also developed a novel methodology for comparing dynamic graph generative models to measure how well they capture: (1) the underlying graph structure distribution, and (2) the node behavior of a real graph over time. In the case of node behavior, using node-aligned metrics over the graph snapshots helps to evaluate the node's topological connectivity and temporal activity. Our use of the Kolmogorov-Smirnov (KS) test with the inter-quartile range, instead of the mean and median values, is an initial effort on adapting graph structure metrics designed for static graphs to the dynamic graph setting. In conclusion, when jointly considering graph structure and node behavior, \modelacronym{} shines in our quantitative evaluation on five different real-world datasets.

\begin{acks}

This research is supported by NSF under contract numbers CCF-0939370 and IIS-1618690.
This work was performed under the auspices of the U.S. Department of Energy by Lawrence Livermore National Laboratory under Contract DE-AC52-07NA27344.
This document was prepared as an account of work sponsored by an agency of the United States government. Neither the United States government nor Lawrence Livermore National Security, LLC, nor any of their employees makes any warranty, expressed or implied, or assumes any legal liability or responsibility for the accuracy, completeness, or usefulness of any information, apparatus, product, or process disclosed, or represents that its use would not infringe privately owned rights. Reference herein to any specific commercial product, process, or service by trade name, trademark, manufacturer, or otherwise does not necessarily constitute or imply its endorsement, recommendation, or favoring by the United States government or Lawrence Livermore National Security, LLC. The views and opinions of authors expressed herein do not necessarily state or reflect those of the United States government or Lawrence Livermore National Security, LLC, and shall not be used for advertising or product endorsement purposes. LLNL-CONF-819670

\end{acks}

\bibliographystyle{ACM-Reference-Format}
\bibliography{main}

\appendix
\section{Appendix} \label{Appendix}

\small
\begin{table}[h!]
    \caption{Statistics of the Dynamic Network Datasets}
\label{tab:dataset_statistics}
    \vspace{-2mm}
    \begin{tabular}{lrrrr}
        \toprule
        Dataset & $|V|$ & $|E|$ & Unique & Timesteps \\
        \midrule
        Enron Emails & 785 & 5,794  & 2,517 & 20 \\
        EU Emails & 784 & 68,091 & 6,878 & 79 \\
        DNC Emails & 1,579 & 33,378 & 3,911 & 23 \\
        Facebook Wall-Posts & 2,245 & 23,507 & 4,976 & 43 \\
        CollegeMsg & 1,083 & 34,328 & 5,589 & 31 \\
        \bottomrule
    \end{tabular}
\end{table}

\begin{figure}
 \centering
\includegraphics[width=0.71\linewidth]{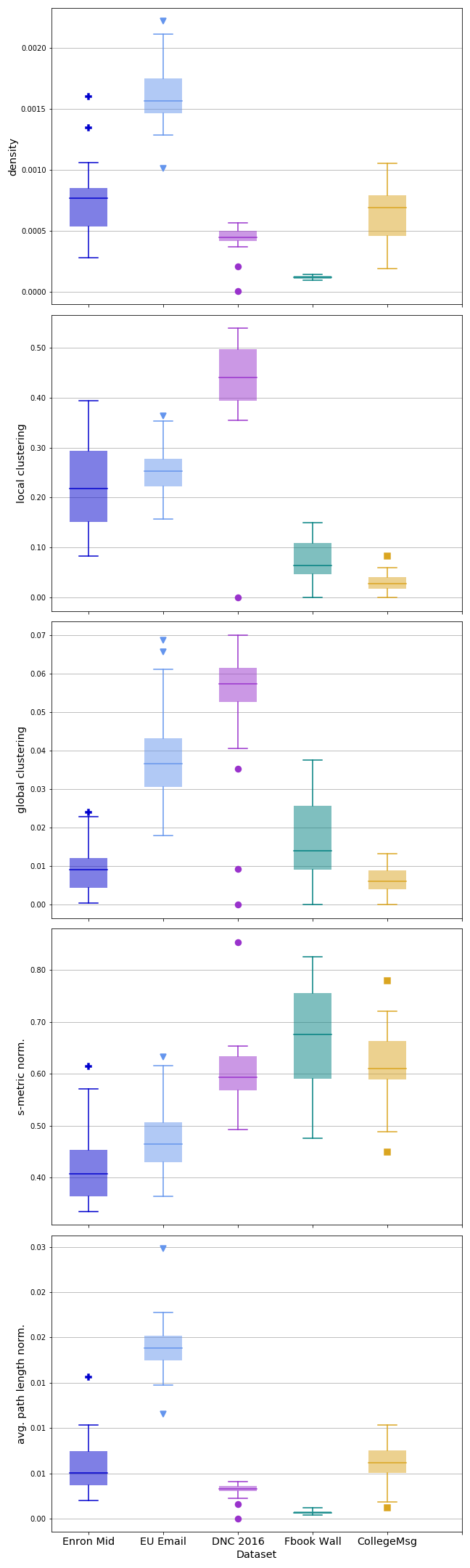}
\caption{Dataset Graph Structure Metrics}
 \label{fig:dataset_graph_metrics}
 \Description{Dataset Graph Structure Metrics}
\end{figure}

\begin{algorithm}[hbt!]
    \SetKwInput{Input}{input}
    \SetKwInOut{Output}{output}
    \DontPrintSemicolon
    \Input{\hspace{2pt}
        $T$, $N$, $\lambda_V$
}
    \Output{$V$ \texttt{// active nodes per timestep}}
    \Begin{
        $V_{t} \leftarrow \emptyset,\ \forall j \in [1, \ldots, T]$ \;
        \For{$v \in [1, \ldots, N]$}{
            $a \sim Exp(\lambda_V)$ \tcp*[h]{sample arrival time} \nllabel{node_active_time} \;
            \For{$t \in [a, \ldots, T]$}{
$V_{t} \leftarrow V_{t} \cup  \{v\}$ \tcp*[h]{add to active nodes} \;
}
        }
        $V = V_{1} \cup \ldots \cup V_{T}$
    }
    \caption{\texttt{GetActiveNodes}}\label{GetActiveNodes}
\end{algorithm}
\begin{algorithm}[hbt!]
    \SetKwInput{Input}{input}
    \SetKwInOut{Output}{output}
    \DontPrintSemicolon
    \Input{\hspace{2pt}
        $t$, $M_{t}$, $M^T$, $\lambda_M = (\lambda_M^{(1)}, \lambda_M^{(2)}, \lambda_M^{(3)})$
}

    \Output{$M^S_{t}$ \texttt{// motif timesteps}}

\Begin{
        \For{$\{u,v,w\} \in M_{t}$}{
$i \leftarrow M^T\big(\{u,v,w\}\big)$ \tcp*[h]{motif type} \;
            \tcp{sample inter-arrival rate}
            $\beta_M\big(\{u,v,w\}\big) \sim Exp\big(\lambda_M^{(i)}\big)$ \nllabel{sample_motif_scale} \;
            $\lambda_M\big(\{u,v,w\}\big) \leftarrow \frac{1}{\beta_M(\{u,v,w\})}$\;
$prev \leftarrow t$ \tcp*[h]{first time motif can appear} \;
            \tcp{sample inter-arrival time}
            $next \sim Exp\big(\lambda_M\big(\{u,v,w\}\big)\big)$ \nllabel{sample_motif_first_time} \;
\While{$prev + next < T$}{
            \tcp{save timestamp to list}
                $M^S_{t}\big(\{u,v,w\}\big)\texttt{.append}(prev + next)$ \;
                $prev \leftarrow prev + next$ \;
                \tcp{sample next inter-arrival time}
                $next \sim Exp\big(\lambda_M\big(\{u,v,w\}\big)\big)$ \nllabel{sample_motif_next_time} \;
            }
}
    }
    \caption{\texttt{SampleMotifTimesteps}}\label{SampleMotifTimesteps}
\end{algorithm}
\begin{algorithm}[hbt!]
\SetKwInput{Input}{input}
     \SetKwInOut{Output}{output}
     \DontPrintSemicolon
     \Input{\hspace{2pt}
         $M_{t}$, $M^T_{t}$, $M^R_{t}$
}
     \Output{$M^E_{t}$ \texttt{// edges for motifs in $M^{(i)}$}}
     \Begin{
         \For{$m \in  M_{t}$}{
             \If(\tcp*[h]{triangle}){$M^T_{t}(m) = 3$}{
                 $M^E_{t}(m) \leftarrow {m \choose 2} $ \;
             }
             \ElseIf(\tcp*[h]{wedge}){$M^T_{t}(m) = 2$}{
                 $h \leftarrow M^R_{t}(m, \texttt{hub})$ \;
                 $s_1, s_2 \leftarrow M^R_{t}(m, \texttt{spoke})$ \;
                 $M^E_{t}(m) \leftarrow \{(h, s_1), (h, s_2)\} $ \;
             }
             \ElseIf(\tcp*[h]{1-edge}){$M^T_{t}(m) = 1$}{
                 $e_1, e_2 \leftarrow M^R_{t}(m, \texttt{equal2})$ \;
                 $M^E_{t}(m) \leftarrow \{(e_1, e_2)\} $ \;
             }
         }
     }
     \caption{\texttt{PlaceMotifEdges}}\label{PlaceMotifEdges}
\end{algorithm}
\begin{algorithm}[hbt!]
     \SetKwInput{Input}{input}
     \SetKwInOut{Output}{output}
     \DontPrintSemicolon
     \Input{\hspace{2pt}
         $M$, $M^S$, $M^E$
}
     \Output{$\textbf{G}' = \{ G'_{1}, \ldots, G'_{T} \}$ \texttt{// where $G'_{t} = (V'_{t}, E'_{t}, S'_{t})$}
     }
\Begin{
         $V'_{t} \leftarrow \emptyset;\ E'_{t} \leftarrow \emptyset;\ S'_{t} \leftarrow \emptyset,\ \forall t \in [1, \ldots, T]$ \;
         \For{$\{u,v,w\} \in M$}{
             $t' \leftarrow  \texttt{min}\left( M^S\big(\{u,v,w\}\big) \right)$ \tcp*[h]{first timestep} \;
             $V'_{t} \leftarrow V'_{t} \cup \{u,v,w\},\ \forall t \in [t', \ldots, T]$
\tcp*[h]{update} \;
             \For{$t \in M^S\big(\{u,v,w\}\big)$}{
$E'_{t} \leftarrow E'_{t} \cup M^E\big(\{u,v,w\}\big)$ \tcp*[h]{place edges} \;
                 \tcp{add timestamps to edges}
                 $S'_{t}\big((u',v')\big) = t,\ \forall (u'v') \in  M^E\big(\{u,v,w\}\big)$ \;
             }
         }
     }
     \caption{\texttt{ConstructGraph}}\label{ConstructGraph}
\end{algorithm}
\begin{algorithm}[hbt!]
    \SetKwInput{Input}{input}
    \SetKwInOut{Output}{output}
    \DontPrintSemicolon
    \Input{\hspace{2pt}
        $\textbf{G} = \{ G_{1}, \ldots, G_{T} \}$ \texttt{// graph to learn from}
    }
    \Output{$\widehat{\lambda}_{V}$ \texttt{// node arrival rate}\\
        $\widehat{\lambda}_M = (\widehat{\lambda}_M^{(1)}, \widehat{\lambda}_M^{(2)}, \widehat{\lambda}_M^{(3)})$ \texttt{// motif rates distr.}\\
        $\widehat{p}_M = (\widehat{p}_M^{(1)}, \widehat{p}_M^{(2)}, \widehat{p}_M^{(3)})$ \texttt{// proportions motifs}\\
        $p_R$ \texttt{// node roles probabilities}\\
        $c_R$ \texttt{// node roles counts}
    }
    \SetKwFunction{GetMotifsGraph}{GetMotifsGraph}
    \SetKwFunction{GetNodeRoleCounts}{GetNodeRoleCounts}
    \Begin{
        Estimate node arrival rate $\widehat{\lambda}_{V}$\hspace{2pt} \tcp*[h]{\autoref{eq:node_arrival_rates}} \;
        $M, M^T \leftarrow $ \GetMotifsGraph{$\textbf{G}$} \;
        \For{$i \in [1,2,3]$}{
            Estimate proportions $\widehat{p}_M^{(i)}$\hspace{2pt} \tcp*[h]{\autoref{eq:motif_proportions}} \;
        }
        \For{$\{u,v,w\} \in M$}{
            Estimate inter-arrival rate $\widehat{\lambda}_M\big(\{u,v,w\}\big)$\hspace{1pt} \tcp*[h]{\autoref{eq:motif_interarrival_rate}} \;
        }
        \For{$i \in [1,2,3]$}{
            Estimate rates of inter-arrivals $\widehat{\lambda}_M^{(i)}$\hspace{2pt}  \tcp*[h]{\autoref{eq:motif_type_interarrival_rates}} \;
        }
        $c_R \leftarrow $ \GetNodeRoleCounts{$M, M^T, M^E, c_R, T$} \;
        \For{$v \in V$}{
            \For{$r \in R$}{
                Estimate role probability $\widehat{p}_R(v, r)$\hspace{2pt} \tcp*[h]{\autoref{eq:node_role_probabilities}} \;
            }
        }
}
    \caption{\texttt{LearnParameters}}\label{LearnParameters}
\end{algorithm}
\begin{algorithm}[hbt!]
\SetKwInput{Input}{input}
    \SetKwInOut{Output}{output}
    \DontPrintSemicolon
    \Input{\hspace{2pt}
        $\textbf{G} = \{ G_{1}, \ldots, G_{T} \}$
}
    \Output{$M\ \ $ \texttt{// motifs in } $\textbf{G}$\\
        $M^T$ \texttt{// motif types}
}
    \Begin{
$M \leftarrow \emptyset$; $M^T \leftarrow \emptyset$ \;
        \For{$t \in [1, \ldots, T]$}{
            \For{$(u,v) \in E_{t}$}{
                \If{$u \neq v$}{
                    \For{$w \in V_{t}$}{
                        \If{$w \neq u$ {\normalfont \textbf{and}} $w \neq v$}{
                            \tcp{type $i$ = num. unique edges}
$i = |E_{t}\big(\{u,v,w\}\big)|$ \;\label{alg:find_motif_type}
                            \If{$\{u,v,w\} \notin M$}{
                                $M \leftarrow M \cup \big\{ \{u,v,w\} \big\}$ \;
                                $M^T\big( \{u,v,w\} \big) \leftarrow i$ \;
}
                            \ElseIf(\tcp*[h]{prefer higher-order motif types}){$i > M^T\big( \{u,v,w\} \big)$}{
                                $M^T\big( \{u,v,w\} \big) \leftarrow i$ \;\label{alg:prefer_higher_order}
}
                        }
                    }
                }
            }
        }
    }
    \caption{\texttt{GetMotifsGraph}}\label{GetMotifsGraph}
\end{algorithm}
\begin{algorithm}[hbt!]
\SetKwInput{Input}{input}
    \SetKwInOut{Output}{output}
    \DontPrintSemicolon
    \Input{\hspace{2pt}
        $M$, $M^T$, $E$, $c_R$, $T$
}
    \Output{$c_R = count(v, r)$ \texttt{// node role counts}}
    \SetKwFunction{GetRoleTimestep}{GetRoleTimestep}
    \Begin{
        $count(v, r) = 0,\ \forall v \in V,\ \forall r \in roles$ \tcp*[h]{init. counts} \;
        \For{$t \in [1, \ldots, T]$}{
            \For{$(u,v) \in E_t$}{
                \If(\tcp*[h]{triangles}){$|N^{(3)}(u,v)| > 0$}{
                    $role\_weight \leftarrow \frac{min(c_t(u,v), |N^{(3)}(u,v)|)}{\frac{|N^{(3)}(u,v)|}{3}}$ \;
                    \For{$n \in [u,v,w]$}{
$count(n, \texttt{equal3}) \pluseq role\_weight$ \;
                    }
                }
\If(\tcp*[f]{wedges (\autoref{eq:remaining_count})}){$|N^{(2)}(u,v)| > 0$ \textbf{ and } $r^{(2)}_t(u,v) > 0$}{
                    $role\_weight \leftarrow \frac{min(r^{(2)}_t(u,v), |N^{(2)}(u,v)|)}{\frac{|N^{(2)}(u,v)|}{2}}$ \;
                    \For{$n \in [u,v,w]$}{
                        $r \leftarrow $ \GetRoleTimestep{$E_t, \{u,v,w\}, n$} \;
                        \If{$r = \texttt{hub}$}{
$count(n, \texttt{hub}) \pluseq role\_weight$ \;
                        }\Else{
$count(n, \texttt{spoke}) \pluseq \frac{role\_weight}{2}$ \;
                        }
                    }
                }
\If(\tcp*[h]{1-edge (\autoref{eq:remaining_count})}){$|N^{(1)}(u,v)| > 0$ \textbf{ and } $r^{(1)}_t(u,v) > 0$}{
                    $role\_weight \leftarrow \frac{min(r^{(1)}_t(u,v), |N^{(1)}(u,v)|)}{|N^{(1)}(u,v)|}$ \;
                    \For{$n \in [u,v,w]$}{
                        $r \leftarrow $ \GetRoleTimestep{$E_t, \{u,v,w\}, n$} \;
                        \If{$r = \texttt{equal2}$}{
$count(n, \texttt{equal2}) \pluseq min(r^{(1)}_t(u,v), |N^{(1)}(u,v)|)$ \;
                        }\Else{
$count(n, \texttt{outlier}) \pluseq role\_weight$ \;
                        }
                    }

                }
            }

        }
    }
    \caption{\texttt{GetNodeRoleCounts}}\label{GetNodeRoleCounts}
\end{algorithm}

 \clearpage
\newpage

\begin{figure}[hb!]
\centering
    \begin{subfigure}[b]{0.40\textwidth}
        \includegraphics[trim={.15cm .15cm .15cm 1.1cm},clip,width=\linewidth]{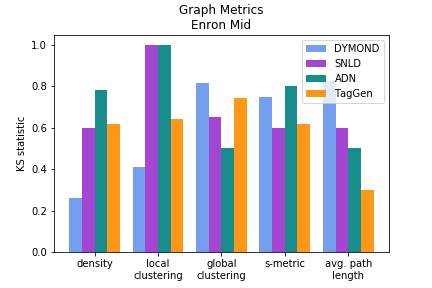}
        \caption{Enron Emails}
        \label{fig:dataset_graph_metrics-Enron}
        \Description{Enron Emails - Graph Structure Metrics}
    \end{subfigure}
    \\
    \begin{subfigure}[b]{0.40\textwidth}
        \includegraphics[trim={.15cm .15cm .15cm 1.1cm},clip,width=\linewidth]{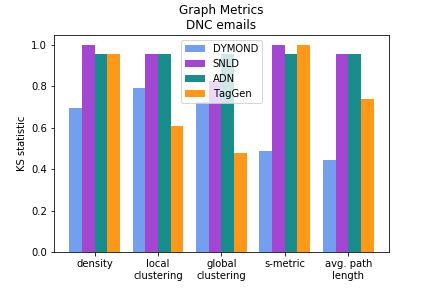}
        \caption{DNC Emails}
        \label{fig:dataset_graph_metrics-DNC}
        \Description{DNC Emails - Graph Structure Metrics}
    \end{subfigure}
    \\
    \begin{subfigure}[b]{0.40\textwidth}
        \includegraphics[trim={.15cm .15cm .15cm 1.1cm},clip,width=\linewidth]{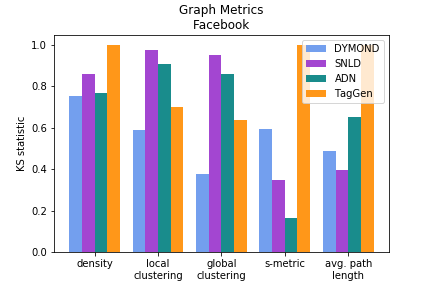}
        \caption{Facebook Wall-Posts}
        \label{fig:dataset_graph_metrics-Facebook}
        \Description{Facebook - Graph Structure Metrics}
    \end{subfigure}
    \\
    \begin{subfigure}[b]{0.40\textwidth}
        \includegraphics[trim={.15cm .15cm .15cm 1.1cm},clip,width=\linewidth]{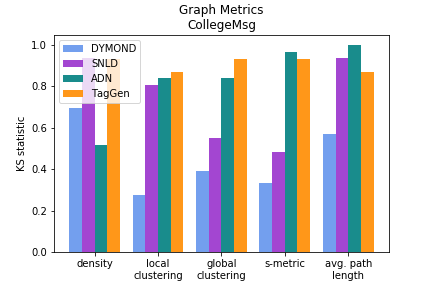}
        \caption{CollegeMsg}
        \label{fig:dataset_graph_metrics-CollegeMsg}
        \Description{CollegeMsg - Graph Structure Metrics}
    \end{subfigure}
\caption{KS Statistic of Graph Structure Metrics}
\end{figure}
\begin{figure}[hb!]
\centering
    \begin{subfigure}[b]{0.40\textwidth}
        \includegraphics[trim={.15cm .15cm .15cm 1.1cm},clip,width=\linewidth]{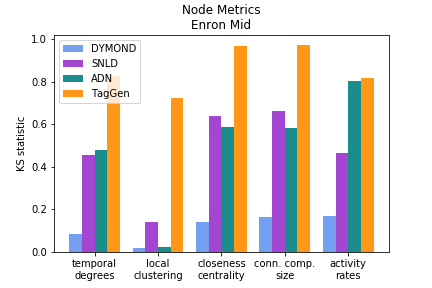}
        \caption{Enron Emails}
        \label{fig:dataset_node_metrics-Enron}
        \Description{Enron Emails - Node Behavior Structure Metrics}
    \end{subfigure}
    \\
    \begin{subfigure}[b]{0.40\textwidth}
        \includegraphics[trim={.15cm .15cm .15cm 1.1cm},clip,width=\linewidth]{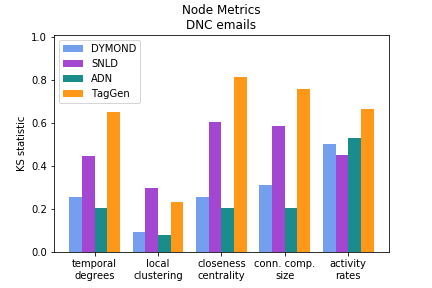}
        \caption{DNC Emails}
        \label{fig:dataset_node_metrics-DNC}
        \Description{DNC Emails - Node Behavior Structure Metrics}
    \end{subfigure}
    \\
    \begin{subfigure}[b]{0.40\textwidth}
        \includegraphics[trim={.15cm .15cm .15cm 1.1cm},clip,width=\linewidth]{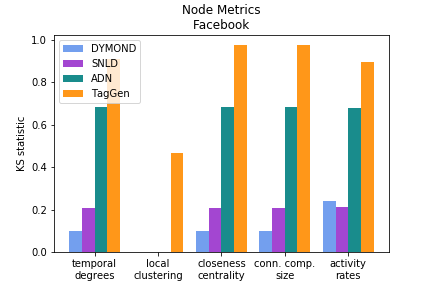}
        \caption{Facebook Wall-Posts}
        \label{fig:dataset_node_metrics-Facebook}
        \Description{Facebook - Node Behavior Structure Metrics}
    \end{subfigure}
    \\
    \begin{subfigure}[b]{0.40\textwidth}
        \includegraphics[trim={.15cm .15cm .15cm 1.1cm},clip,width=\linewidth]{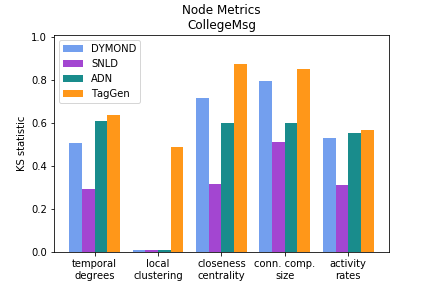}
        \caption{CollegeMsg}
        \label{fig:dataset_node_metrics-CollegeMsg}
        \Description{CollegeMsg - Node Behavior Structure Metrics}
    \end{subfigure}
\caption{KS Statistic of Node Behavior Structure Metrics}
\end{figure}

\end{document}